\documentstyle[12pt]{article}
\newcommand{\be}{\begin{equation}}
\newcommand{\ee}{\end{equation}}
\begin{document}
\begin{titlepage}
\title{
{\bf Interaction Hierarchy.\\
Gonihedric String and Quantum Gravity
}
}
{\bf
\author{
 G.K. Savvidy\\
 Physics Department, University of Crete \\
 71409 Iraklion, Crete, Greece ; \\
 Yerevan Physics Institute, 375036 Yerevan, Armenia\\
  \vspace{1cm}\\
  K.G. Savvidy\\
 Princeton University, Department of Physics\\
 P.O.Box 708, Princeton, New Jersey 08544, USA
}
}
\date{}
\maketitle
\begin{abstract}
\noindent
We have found that the Regge gravity
\cite{regge,sorkin},
can be represented as a $superposition$ of less
complicated theory of random surfaces with $Euler~character$ as an
action. This extends  to Regge gravity
our previous result \cite{savvidy}, which allows to
represent the gonihedric string \cite{savvidy1}
as a superposition of less complicated theory of
random paths with $curvature$ action. We propose also an
alternative linear action $A(M_{4})$ for the four and high dimensional
quantum gravity.
{}From these representations it follows that the corresponding
partition functions are equal to the product of Feynman path integrals
evaluated on time slices with curvature and length action for the
gonihedric string and with Euler character and
gonihedric action for
the Regge gravity. In both cases the interaction
is proportional to the overlapping
sizes of the paths or surfaces on the neighboring time slices. On the
lattice we constructed spin system with local interaction, which have
the same partition function as the quantum gravity. The scaling limit is
discussed.
\end{abstract}
\thispagestyle{empty}
\end{titlepage}
\pagestyle{empty}

{\Large{\bf 1.}} In our previous article \cite{savvidy} ,
devoted to the hierarchical structure
of the geometrical interactions, we observed that the physical theories
can be considered as a $superposition$ of less complicated, primary
interactions.

The example on which we have been based is the regularized string theory
with linear action $A(M_2)$ \cite{savvidy1}.
This string can be considered as a superposition of less complicated
theory of random paths $\{M_1\}$ with an amplitude which is
proportional to the total curvature of the path $k(M_1)$ \cite{schneider}.

In the present article we will extend this result  and will show that
the regularized quantum gravity \cite{regge,sorkin,gross1,gross2,adler}
can be represented as a superposition
of the random surfaces $\{M_2\}$ with an action  which is equal to the
Euler character of the surfaces $\chi(M_2)$. We propose an
alternative linear action $A(M_{4})$ for the four and higher
dimensional quantum gravity.

In the next section we present the main ideas and
results for the regularized
string with linear action $A(M_2)$ and in the subsequent sections we will
extend this result to the quantum gravity.

{\Large{\bf 1.1}}The regularized string with linear-gonihedric
action $A(M_2)$
can be derived from natural
physical requirements such as \cite{savvidy1} :
$\alpha )$ coincidence of the string transition amplitude
with the usual Feynman path amplitude for long space-time strips
and from $\beta )$ the continuity principle for the transition amplitudes.

The partition function of the regularized linear string
has the form \cite{savvidy1}
\be
Z_{Gonihedric}(\beta) = \sum_{\{ M_2 \}}
\exp\{- \beta A(M_2) \},~~~~~~~~~A(M_2) = \sum_{<i,j>}  \lambda_{i,j}
\cdot \Theta(\alpha_{ij}) \label{Z},
\ee
where~ $\Theta(\alpha_{ij})=\vert \pi - \alpha_{i,j} \vert $~and
the summation is extended over all triangulated
surfaces $\{ M_2 \}$ with the linear action $A(M_2)$,
and $\alpha_{i,j}$~ is the dihedral angle between two neighbor
faces of $M_2$ having a common edge $<i,j>$ of the length
$\lambda_{i,j}$. The regularized string (\ref{Z})
is well defined in any dimensions and
for an arbitrary topology of the surface $M_2$.

The linear string (\ref{Z}) can be viewed as a superposition of
less complicated, primary theory of random paths $\{M_1\}$
with the curvature action $k(M_1)$.
This  structure of the linear string (\ref{Z}) comes from
the representation of the linear
action $A(M_2)$ in terms of the total curvature $k(M^{E}_{1})$ of the
paths $\{ M^{E}_{1} \}$ which appear in the
intersection of the plane $E$ with
the given surface $M_2$ \cite{schneider}

\be
 A(M_2) = \sum_{\{ E \} }~k(M^{E}_{1}) \label{cur}.
\ee
In the last formula the paths in the intersection are denoted
by $\{ M^{E}_{1} \}$

\be
M^{E}_{1} = M_{2} \cap  E
\ee
and the absolute total curvature $k(M^{E}_{1})$ of the path $M^{E}_{1}$
is equal to

\be
 k(M^{E}_{1}) = \sum_{<i,j>} \vert \pi-\alpha^{E}_{ij} \vert ,
\ee
where $\alpha^{E}_{ij}$ is the dihedral angle in the intersection
of the plane $E$ with the edge $<i,j>$ and $\alpha^{E}_{ij}=\pi$
for the edges of $M_2$ which are not intersected by the given
plane $E$.

With (\ref{cur}) the partition function of the
system (\ref{Z}) can be written in the form

\be
Z_{Gonihedric}(\beta)
= \sum_{\{M_2\}}~ \prod_{\{ E \}} exp\{-\beta~k(M^{E}_{1})~\}.
\ee
When the continuous Euclidean space is replaced by the Euclidean
lattice, where the paths and the surfaces are associated with
the collection of the links and plaquettes, then
in the last formula the product over all intersecting planes
$\{E\}$ can be evaluated
to a product over planes $\{E^{\tau}\}$ which are perpendicular
to a given time direction $\tau$ \cite{savvidy}

\be
Z_{Gonihedric}(\beta) =
\sum_{\{..M^{\tau}_{1},M^{\tau+1}_{1}..\}}~ \prod_{\tau}~K(M^{\tau}_{1},
M^{\tau+1}_{1})
\label{Dirac},
\ee
where

\be
K(M^{\tau}_{1},M^{\tau +1}_{1}) =
exp -\beta \{ \frac{1}{2} k(M^{\tau}_{1}) + l(M^{\tau}_{1})
+\frac{1}{2} k(M^{\tau+1}_{1}) + l(M^{\tau +1}_{1})
-2 l(M^{\tau}_{1} \cap M^{\tau +1}_{1})~ \} \label{Feynman}
\ee
and the $independent$ summation is extended over all paths
$\{..M^{\tau}_{1},M^{\tau+1}_{1}..\}$ on diferent time slices.
This result is valid if the self-intersection
coupling constant $k$ \cite{savvidy1} is equal
to infinity \cite{savvidy2}, that is for
$self-avoiding$ surfaces. In three
dimensions it is valid also for the case when $k=0$
\cite{savvidy}.

We have the propagation of the path $M^{\tau}_{1}$ in the
time direction $\tau$
with an amplitude which is proportional to the sum of the curvature
$k(M^{\tau}_1)$ and of the length of the path $l(M^{\tau}_{1})$,
and the interaction which is proportional to the
overlapping length of the paths on the neighboring time slices
$l(M^{\tau}_{1} \cap M^{\tau+1}_{1})$. The advantage of this
formula is that one can consider this interaction as a perturbation,
or as the $hopping$ term in the lattice language,
because it describes the hopping from one time slice to another
\cite{savvidy}.
In that case the tree approximation is associated
with the length and curvature terms in (\ref{Feynman}) and
describes the $free~fermion$ on a given time slice.
The separate path integral which describes the fermion on a given time
slice

\be
\sum_{\{ M^{\tau}_{1}\} }
exp\{-\beta~ \{ \frac{1}{2}~ k(M^{\tau}_{1}) +
l(M^{\tau}_{1}) \}~\} = <exp\{ - H^{tree}_{Gonihedric}\}>
\ee
has been already computed in \cite{savvidy} by Kac-Ward method
\cite{waerden,polyakov}.

 From the last formulas one can conclude that the string
partition function is equal to the limit of the
infinite product of the Dirac-Feynman path integrals which are
evaluated on the time slices $\{E^{\tau}\}$~~$\tau = a,1,...,N,b$.
One can say that the gonihedric string (\ref{Z}) is a superposition of
weakly interacting Dirac particles which "jumps" from one time slice
to another thanks to the hopping interaction \cite{savvidy}.
This result demonstrates how the $infinite~sandwich$ of primary
theories can generate the physical theory.

The equivalent statistical  system
for which (\ref{Dirac}),(\ref{Feynman})
is an exact expression for the partition function has been
constructed in \cite{wegner,savvidy2,wegner1}.
This equivalence allows to simulate the
linear string (\ref{Z}) on the lattice and, as we will see later on,
the quantum gravity as well.

{\Large{\bf 2.}}Let us consider three dimensional manifold $\{M_3\}$
which is constructed  by
gluing together three-dimensional tetrahedra through their triangular
faces. This manifold can be associated with the three-dimensional
gravity \cite{regge,sorkin}
or with the motion of the two-dimensional membrane \cite{savvidym}.

To construct the three dimensional quantum gravity with the linear action
$A(M_3)$ and to generate the next species in the hierarchy of the
geometrical interactions we should apply the
principles $\alpha )$ and $\beta )$.
In accordance with $\alpha )$ the quantum mechanical amplitude should be
proportional to the linear size of the manifold and thus  it must be
proportional to the linear combination of the lengths of all edges of
tetrahedrated manifold $M_3$

\be
A(M_3) = \sum_{<i,j>} \lambda_{ij}\cdot \Theta_{ij}
\ee
where $\lambda_{ij}$ is the length of the edge between
two vertices $<i>$ and $<j>$, summation is over all
edges $<i,j>$  and
$\Theta_{ij}$ is unknown factor, which can be defined by use of
the continuity principle $\beta )$. Indeed, if we impose a new
vertex $<m>$ inside a given $flat$ tetrahedron $<ijkl>$, then for that
new manifold we will get an extra contribution
$\lambda_{im}\Theta_{im}+\lambda_{jm}\Theta_{jm}+\lambda_{km}\Theta_{km}+
\lambda_{lm}\Theta_{lm}$ to the action and we will get more extra
terms imposing a new vertices, despite the fact that the manifold
has not actually changed. To exclude such type of contributions we
should choose unknown factor $\Theta_{ij}$ such that it will vanish
in $flat$ cases. This can be done by use of the dihedral angles,
therefore
\be
A(M_3) = \sum_{<i,j>} \lambda_{ij}\cdot
\Theta(\alpha_{ij}+\beta_{ij}+...) , \label{Hilquan}
\ee
where

\be
\Theta(2\pi) = 0,
\ee
and $\alpha_{ij}, \beta_{ij}...$ are the dihedral angles between
triangular faces of tetrahedra which have the common edge $<i,j>$.
These angles appear in the normal
section of the edge $<i,j>$ with the $d-1$ dimensional plane $E$.
In analogy with the gonihedric string \cite{savvidy1}
one can also define more specific theories with the property that

\be
\Theta(4\pi-\omega) = \Theta(\omega),~~~~~~~~~~~~\Theta(\omega) \geq  0
\ee
As we will see bellow this generalization will allow
to extend the string results to quantum gravity.

The essential difference with the theory of random surfaces with the
linear action $A(M_2)$
is that now the linear functional $A(M_3)$ is an $intrinsic$ quantity,
because dihedral angles $\alpha_{ij}, \beta_{ij}...$ are well defined
without any embedding.

Generally speaking, a suitable selection of the factor $\Theta$ can be done
if we require a convenient scaling behavior of the theory \cite{savvidy1}.
We will use the following parametrization of the $\Theta(\omega)$

\be
\Theta(\omega)= ( 2\pi - \omega )^{\varsigma},
\ee
which for the case $\varsigma = 1$ coincides with the
Regge action \cite{regge}
and is the discrete version of the following
continuous Hilbert-Einstein
action

\be
A(M_{3})= \int_{M_3}R~ dv_3 . \label{Hil}
\ee
Our aim is now to represent the three dimensional gravity as a
$superposition$ of less complicated geometrical theory
of random surfaces with Euler character as an action

\be
\chi(M_2) = \sum_{<i,j>} (2\pi - \alpha_{ij}- \beta_{ij}-... ).
\label{char}
\ee
In the last particular case $\varsigma = 1$ and $\Theta \equiv \chi$.

Let us consider the intersection of the tetrahedrated manifold
$M_3$  by the $d-1$ dimensional plane $E$. The intersection is
the two-dimensional surface $M^{E}_{2}$

\be
M^{E}_{2} = M_{3} \cap E
\ee
and as we will see in a moment the linear action $A(M_3)$ is
the sum of
Euler characters of all surfaces which appear in the intersection
$\{ M^{E}_{2}\}$

\be
A(M_3) = \sum_{\{ E\}}~\chi(M^{E}_{2}~) , \label{super}
\ee
where

\be
\chi(M^{E}_{2}) = \sum_{<i,j>} (2\pi -
\alpha^{E}_{ij}- \beta^{E}_{ij}-... ), \label{Eulercha}
\ee
and $\alpha^{E}_{ij}$, $\beta^{E}_{ij}$, ... are the
angles in the intersection of the plane $E$ with the edge $<i,j>$
and $\omega^{E}_{ij}= \alpha^{E}_{ij} +
\beta^{E}_{ij}+... = 2\pi$ for the edges of $M_3$ which are not
intersected by the given plane $E$.

The formula (\ref{super}) follows from the fact that
the average of the angle $\alpha^{E}_{ij}$ over
all intersecting planes $\{ E \}$ is equal to dihedral angle $\alpha_{ij}$
\cite{schneider}

\be
<\alpha^{E}_{ij}> \equiv \int \alpha^{E}_{ij}~ dE = \alpha_{ij}
\cdot \lambda_{ij}
\ee
and that the number of planes which intersect the given edge $<i,j>$
is proportional to it's length $\lambda_{ij}$ .

Therefore in the same way as for the linear string $A(M_2)$ we
have the following representation of the partition function
of the three dimensional quantum gravity

\be
Z_{Gravity}(\beta)
= \sum_{\{M_3\}}~ \prod_{\{ E \}}~ exp\{-\beta~\chi(M^{E}_{2})~\}.
\label{statsuper}
\ee
When the continuous Euclidean space is replaced by the Euclidean
lattice, where the surfaces and the manifolds are associated with
the collection of the plaquettes and cubes, then
in the last formula the product over all intersecting planes
$\{E\}$ can be evaluated
to a product over planes $\{E^{\tau}\}$ which are perpendicular
to a given time direction $\tau$

\be
Z_{Gravity}(\beta) =
\sum_{\{..M^{\tau}_{2},M^{\tau+1}_{2}..\}}~ \prod_{\tau}~K(M^{\tau}_{2},
M^{\tau+1}_{2})
\label{DiGrav},
\ee
where

\be
K(M^{\tau}_{2},M^{\tau +1}_{2}) =
exp -\beta \{ \frac{1}{2} \Theta(M^{\tau}_{2}) + A(M^{\tau}_{2})
+\frac{1}{2} \Theta(M^{\tau+1}_{2}) + A(M^{\tau +1}_{2})
-2 A(M^{\tau}_{2} \cap M^{\tau +1}_{2})~ \} \label{FeyGrav}
\ee
and the $independent$ summation is extended
over all surfaces $\{.. M^{\tau}_{2},M^{\tau+1}_{2}.. \}$ on
different time slices.
This formula is valid when
\be
\Theta(\omega)=\vert 2\pi - \omega \vert \label{Eulerchag}
\ee
and self-intersection coupling constant $k$ is equal to
infinity \cite{savvidy2}.
For three-dimensional universe $M_3$ self-intersection
happens when more than two 3d cubes have been glued face-to-face.
To prove this formula one should
consider case by case all seven vertex configuration
$\theta_{1},...,\theta_{7}$ which remain in the limit
$k \rightarrow \infty$ (see bellow).

We have the propagation of the surface $M^{\tau}_{2}$ in the
time direction $\tau$
with an amplitude which is proportional to the sum of the
generalized Euler
character $\Theta(M^{\tau}_{2})$ (\ref{Eulerchag}) and
of the linear size of the surface
$A(M^{\tau}_{2})$~~ (\ref{Z}),
and the interaction which is proportional to the length of the
right angle edges of the
overlapping  surface $A(M^{\tau}_{2} \cap M^{\tau+1}_{2})$.

 From the last formulas we conclude that the  partition
function for the quantum gravity is equal to the limit of the
infinite product of the linear string path integrals
(\ref{DiGrav}),(\ref{FeyGrav}) which are
evaluated on every time slices $\tau = a,1,...,N,b$.

{\Large{\bf 2.1}} Let us consider now four dimensional manifold
$M_{4}$. Using the same principles we can define the linear action $A(M_{4})$
for the four dimensinal quantum gravity as

\be
A(M_{4}) =\sum_{<i,j>}\lambda_{ij}\cdot \Theta\{\sum_{q} (2\pi -
\alpha^{q}_{ij} - \beta^{q}_{ij}-...)\}
\ee
where the first summation is extended over all edges of $M_{4}$ and the
second summation is over all $d-2$ dimensional normal sections
of the given edge $<i,j>$, and $\alpha^{q}_{ij},~\beta^{q}_{ij},...$
are dihedral angles on the $q'th$ section. Geometricaly the last
factor is equal to the total area of the polyhedron on $S^{3}$
which corresponds to the spherical image of the edge $<i,j>$.
This linear theory is again $intrinsic$
and have better chances to describe quantum gravity.

For the standart action \cite{regge}, which is proportional to the area
$S(M_{4})$ of the four dimensional universe $M_{4}$
the same arguments as in the previous section allow to find out that

\be
S(M_{4}) = \sum_{\{ E\}}~\chi(M^{E}_{2}~),
\ee
where the summation is extended over all $d-2$ dimensiona planes
$\{E\}$. With this result we have the same representation
(\ref{statsuper}) for the
partition function of the four dimensional Quantum Gravity.
In the next section we will construct the spin systems which
have the equivalent partition functions.

{\Large{\bf 3.}} The correspondence between the
spin configurations and
the geometry of interface allows to define different theories of
random manifolds on a lattice \cite{wegnermathphys,wegner,waerden}.
In the recent articles  \cite{wegner,savvidy2,wegner1}
the authors have introduced a spin
statistical system on the lattice, the $low$ and $high$ temperature
expansion of which generates random walks $\{ M_1 \}$,  with
the amplitude which is proportional
to a total curvature $k(M_1)$ of
the paths on the two-dimensional lattice
and to the linear size of the surface $A(M_2)$ (\ref{Z})
on  three-dimensional lattice. These two spin systems are an
example of the primary theory and of the physical theory which is  an
infinite superposition of the first one,
and allows therefore the reach phase structure \cite{savvidy}.
In three dimensions the corresponding Hamiltonian is
\cite{wegner,savvidy2,wegner1}
$$
H_{gonihedric}^{3d}(k)= k \cdot H_{self-intersections}^{3d} +
H_{gonihedric}^{3d}(0)
$$
\be
=- 2k \sum_{\vec{r},\vec{\alpha}} \sigma_{\vec{r}}
\sigma_{\vec{r}+\vec{\alpha}}
+ \frac{k}{2} \sum_{\vec{r},\vec{\alpha},\vec{\beta}} \sigma_{\vec{r}}
\sigma_{\vec{r}+\vec{\alpha} +\vec{\beta}}
-  \frac{1-k}{2} \sum_{\vec{r},\vec{\alpha},\vec{\beta}} \sigma_{\vec{r}}
\sigma_{\vec{r}+\vec{\alpha}} \sigma_{\vec{r}+\vec{\alpha}+\vec{\beta}}
\sigma_{\vec{r}+\vec{\beta}}, \label{3dhamil}
\ee
and the low temperature expansion of the partition function
is equal to:
\be
Z(\beta)=\sum_{ \{ \sigma \}}
exp( - \beta H_{gonihedric}^{3d} )~~~~~~~~~~= \sum_{ \{ M_2 \}}
exp( - 2\beta A(M_{2}) )
\ee
In this lattice implementation
of the linear string $A(M_2)$ the corresponding Hamiltonian
(\ref{3dhamil}) depends
on the self-intersection coupling constant $k$ and the system simplifies
in the supersymmetric point where the
self-intersection coupling constant is equal to zero $k=0$
\cite{savvidy2,wegner1}
\be
H_{gonihedric}^{3d}(0)=- \frac{1}{2} \sum_{\vec{r},\vec{\alpha}, \vec{\beta}}
\sigma_{\vec{r}}
\sigma_{\vec{r}+\vec{\alpha}}
\sigma_{\vec{r}+\vec{\alpha}+\vec{\beta}}
\sigma_{\vec{r}+\vec{\beta}} \label{susy} .
\ee
The system (\ref{susy}) is highly symmetric, because
one can independently flip spins on any combination of
planes (spin layers) of the lattice ${N^{d}}$.
The degeneracy of the vacuum state is equal to $2^{dN}$
and allows to construct the
dual Hamiltonian \cite{savvidy,pavel,wegner1}.
String path integral (\ref{Dirac}),(\ref{Feynman})
which we discussed in
the previous section is an exact expression for the spin system
with the Hamiltonian (\ref{susy}) and allows
to predict the second order phase transition in 3d which should
be of the same nature as it is in the case of the 2d Ising ferromagnet
\cite{savvidy}. This system has well separated vacuum states
with nonzero generalized
magnetization at low temperature and symmetric state at high temperature.

{\Large{\bf 3.1}}In this section our aim is to
construct the lattice spin system
with the low temperature partition function which is equal to
the partition function of the random surfaces with Euler character.
This will allow to construct the spin system which simulates quantum
gravity with the linear action $A(M_3)$ and $A(M_{4})$ on the lattice.

The Hamiltonian which contains eight different types of spin interactions
inside the $3d$ cube can be written in the form

$$
H^{3d} = a\cdot \sum_{\vec{r},\vec{\alpha},\vec{\beta},\vec{\gamma}}
\sigma_{\vec{r}}\sigma_{\vec{r}+\vec{\alpha}}
\sigma_{\vec{r}+\vec{\alpha}+\vec{\beta}}
\sigma_{\vec{r}+\vec{\beta}}
\sigma_{\vec{r}+\vec{\gamma}}\sigma_{\vec{r}+\vec{\gamma}+\vec{\alpha}}
\sigma_{\vec{r}+\vec{\gamma}+\vec{\alpha}+\vec{\beta}}
\sigma_{\vec{r}+\vec{\gamma}+\vec{\beta}}
$$
$$
+g\cdot \sum_{\vec{r},\vec{\alpha},\vec{\beta},\vec{\gamma}}
\sigma_{\vec{r}}\sigma_{\vec{r}+\vec{\gamma}}
\sigma_{\vec{r}+\vec{\gamma}+\vec{\alpha}}
\sigma_{\vec{r}+\vec{\gamma}+\vec{\alpha}+\vec{\beta}}
\sigma_{\vec{r}+\vec{\alpha}+\vec{\beta}}
\sigma_{\vec{r}+\vec{\beta}}
$$
$$
+b\cdot \sum_{\vec{r},\vec{\alpha},\vec{\beta},\vec{\gamma}}
\sigma_{\vec{r}}\sigma_{\vec{r}+\vec{\gamma}}
\sigma_{\vec{r}+\vec{\alpha}+\vec{\beta}}
\sigma_{\vec{r}+\vec{\gamma}+\vec{\alpha}+\vec{\beta}}
+2c\cdot \sum_{\vec{r},\vec{\alpha},\vec{\beta}}
\sigma_{\vec{r}}\sigma_{\vec{r}+\vec{\alpha}}
\sigma_{\vec{r}+\vec{\alpha}+\vec{\beta}}
\sigma_{\vec{r}+\vec{\beta}}
+e\cdot \sum_{\vec{r},\vec{\alpha},\vec{\beta},\vec{\gamma}}
\sigma_{\vec{r}}\sigma_{\vec{r}+\vec{\alpha}}
\sigma_{\vec{r}+\vec{\beta}}
\sigma_{\vec{r}+\vec{\gamma}}
$$
\be
2d \cdot \sum_{\vec{r},\vec{\alpha},\vec{\beta}}
\sigma_{\vec{r}}\sigma_{\vec{r}+\vec{\alpha}+\vec{\beta}}
+h \cdot \sum_{\vec{r},\vec{\alpha},\vec{\beta},\vec{\gamma}}
\sigma_{\vec{r}}\sigma_{\vec{r}+\vec{\alpha}+\vec{\beta}+\vec{\gamma}}
+4f \cdot \sum_{\vec{r},\vec{\alpha}}
\sigma_{\vec{r}}\sigma_{\vec{r}+\vec{\alpha}}
\label{abcdf}
\ee
where the coupling constants $a,g,b,c,e,d,h$ and $f$ describe eight spin,
six spin (without main diagonal), four diagonal spin, four spin (plaquette),
four spin (around cube vertex), two diagonal spin,
two spin (main diagonal),   and
usual direct two spin interactions terms. It is convenient to
consider the part of the Hamiltonian which belongs to a given
3d cube
$$
H^{3d}_{cube} = a\cdot \sigma\sigma\sigma\sigma\sigma\sigma\sigma\sigma
+g \cdot \sigma\sigma\sigma\sigma\sigma\sigma(four~zigzag~terms)
$$
$$
+b\cdot \sigma\sigma\sigma\sigma(six~diagonal)
+c\cdot \sigma\sigma\sigma\sigma(six~square)
+e\cdot \sigma\sigma\sigma\sigma(eight~terms~around~vertex)
$$
\be
+d\cdot \sigma\sigma(twelve~diagonal)
+h\cdot \sigma\sigma(four~main~diagonal~terms)
+f\cdot \sigma\sigma(twelve~direct).
\ee
There are thirteen
different configurations of the interface in 3d cube first seven of
which are
shown on Fig.1. The corresponding seven basic vertex curvature
are equal to
$$
\theta_{1} = -a +6d +6f+2h-2g,~~~\theta_{2}=a-2b+2c+4f,~~~\theta_{3}=
-a-2d+2f-2h+2g,$$
$$\theta_{4}= a+6b+6c-4d+4f-8e-4h-4g,~~~\theta_{5}= a+6b-6c+8e-4h-
4g,~~~\theta_{6}= a-2b-2c-4d,$$
\be
\theta_{7}= a+6b-6c-8e+4h+4g,
\ee
and six vertices with self-intersections are equal to
$\theta_{8}= a-2b-2c+4d, ~~~\theta_{9}= a-2b+2c-4f,~~~$$\theta_{10}=
-a-2d-2f+2h-2g,~~~\theta_{11}= -a+6d-6f-2h+2g,~~~\theta_{12}=
a+6b+6c-4d-4f+8e+4h+4g,~~~\theta_{13}= a+6b+6c+12d-12f-8e-4h-4g.$
To implement generalized Euler action,
the number of constraints should be imposed on
these weights.

The first requirement is that $\theta_{2}= \theta_{4}=0$,
because this vertices are $flat$,
and one can always normalize $\theta_{1}=1$
and to parametrize $\theta_{3}$ by $\rho$, therefore
$\theta_{1}= 1,~~~\theta_{2}= 0,~~~\theta_{3}=\rho, ~~~\theta_{4}= 0.$
 From this equations we can find the coupling constants $a,b,c$ and $d$
in terms of free parameters $\rho$ and coupling constants $f,g,h,e$.
The solution is:
$a=(72f+24g-24h-18\rho -6)/24,~~~b=(24f+16g+16e-4\rho)/24,~~~c=
(-60f+4g+16e+12h+5\rho +3)/24,~~~d=(-12f+12g-12h-3\rho+3)/24$,
therefore the basic vertex curvature become equal to :
$$
\theta_{1} = 1,~~~\theta_{2}=0,~~~\theta_{3}=\rho,~~~\theta_{4}= 0,$$
$$\theta_{5}= -1-3\rho+24f+8e-8h,~~~\theta_{6}=
-1-\frac{1}{3}\rho +8f-\frac{8}{3}g-\frac{8}{3}e,~~~\theta_{7}=
-1-3\rho +24f+8g-8e,$$
and the vertices with self-intersections are equal to:
$\theta_{8}= -\frac{4}{3}\rho +4f+\frac{4}{3}g-4h-\frac{8}{3}e,~~~\theta_{9}=
-8f,~~~\theta_{10}= \rho-4f-4g+4h,~~~\theta_{11}= -12f+4g-4h+1,~~~\theta_{12}=
-8f+8g+16e+8h,~~~\theta_{13}=2-2\rho -24f+8g-8h$.
The second requirement which should be imposed on the vertex curvature
is that $\theta_{5} =\theta_{6}$ from which it follows that
$6f=\rho+3h-g-4e$.
The coupling constants are parametrized now in terms of $\rho$ and  $g$,
$h$, $e$ and are equal to
$a=(-1-\rho+2g+2h-8e)/4,~~~b=(g+h)/2,~~~c=(
3-5\rho+14g-18h+56e)/24,~~~d=(3-5\rho+14g-18h+8e)/
24,~~~f=(\rho-g+3h-4e)/6$.
Therefore $\theta_{1} = 1,~~~\theta_{2}=0,~~~\theta_{3}=
\rho,~~~\theta_{4}= 0,~~~\theta_{5}= \theta_{6}=
-1+\rho+4h-4g-8e,~~~\theta_{7}=-1+\rho+12h+4g-24e$
and $\theta_{8}= (-2\rho+2g-6h-16e)/3,~~~\theta_{9}=
(-4\rho+4g-12h+16e)/3,~~~\theta_{10}
= (\rho-10g+6h+8e)/3,~~~\theta_{11}=
(-2\rho+6g-10h+8e+1)/3,~~~\theta_{12}=
(-4\rho+28g+12h+64e)/3,~~~\theta_{13}=2-6\rho+12g-20h+16e$.
 From the last condition $\theta_{5} =\theta_{6}=\theta_{7}$
it follows that $2e=g+h$ and

$$a=\frac{-1-\rho-2g-2h}{4},~~~b=
\frac{g+h}{2},~~~c=\frac{3-5\rho+42g+10h}{24},$$
\be~~~d=
\frac{3-5\rho+18g-14h}{24},~~~f=\frac{\rho-3g+h}{6}~~~e=\frac{g+h}{2}
\label{coupling}
\ee
and finally
\be
\theta_{1} = 1,~~~\theta_{2}=0,~~~\theta_{3}=\rho,~~~\theta_{4}=
0,~~~\theta_{5}= \theta_{6}=\theta_{7}= -1+\rho-8g
\label{weight}
\ee
together with self-intersection vertices $\theta_{8}=
(-2\rho-6g-14h)/3,~~~\theta_{9}=
(-4\rho+12g-4h)/3,~~~\theta_{10}
=( \rho-6g+10h)/3,~~~\theta_{11}=
(-2\rho+10g-6h+1)/3,~~~\theta_{12}=
(-4\rho+60g+44h)/3,~~~\theta_{13}=2-6\rho+20g-12h$.
The (\ref{coupling}) and (\ref{weight}) completely solve the
problem in terms of parameter $\rho$ and coupling constants
$g$ and $h$. Bellow we will consider two different cases of
the prime theories.

{\Large{\bf 3.2}} Canonical weights for vertex curvature.
In this case we should take $\rho = -1$ and $g=h=0$ in (\ref{weight}),
then first six vertices
have canonical Euler value
\be
\chi_{1} = 1,~~~\chi_{2}=0,~~~\chi_{3}
=-1,~~~\chi_{4}= 0,~~~\chi_{5}= \chi_{6}=\chi_{7}= -2
\ee
and from (\ref{coupling}), (\ref{abcdf}) the Haliltonian is equal to
\be
H^{3d}_{Euler}=
-\sum_{\vec{r},\vec{\alpha}} \sigma_{\vec{r}}
\sigma_{\vec{r}+\vec{\alpha}}
+ \sum_{\vec{r},\vec{\alpha},\vec{\beta}} \sigma_{\vec{r}}
\sigma_{\vec{r}+\vec{\alpha} +\vec{\beta}}
+ \sum_{\vec{r},\vec{\alpha},\vec{\beta}} \sigma_{\vec{r}}
\sigma_{\vec{r}+\vec{\alpha}} \sigma_{\vec{r}+\vec{\alpha}+\vec{\beta}}
\sigma_{\vec{r}+\vec{\beta}}, \label{Euler}
\ee
The remaining surface vertices
$\chi_{8},..,\chi_{13}$
have the lines of self-intersections and
are equal to $\chi_{8}= \frac{2}{3},~~~\chi_{9}=
\frac{4}{3},~~~\chi_{10}= -\frac{1}{3},~~~\chi_{11}=
3,~~~\chi_{12}=\frac{4}{3},~~~\chi_{13}=8.$
Most of these vertices are positive and therefore have less
statistical weight compared with other basic vertices
$\chi_{1},..,\chi_{7}$.
To exclude them completely from the partition function one can
ascribe to them a large or infinite curvature. For that we
can use the fact that they all have the lines of self-intersections.
The Hamiltonian which counts the number of self-intersection lines
with the weight equal to $k$ has been already constructed
\cite{savvidy2} and is equal to $H^{3d}_{self-intersection}$
(\ref{3dhamil}).
The total Hamiltonian is equal therefore to
\be
H^{3d}_{Euler}=
-(12k+4)\cdot \sum_{\vec{r},\vec{\alpha}} \sigma_{\vec{r}}
\sigma_{\vec{r}+\vec{\alpha}}
+(3k+4)\cdot \sum_{\vec{r},\vec{\alpha},\vec{\beta}} \sigma_{\vec{r}}
\sigma_{\vec{r}+\vec{\alpha} +\vec{\beta}}
+(3k+4)\cdot \sum_{\vec{r},\vec{\alpha},\vec{\beta}} \sigma_{\vec{r}}
\sigma_{\vec{r}+\vec{\alpha}} \sigma_{\vec{r}+\vec{\alpha}+\vec{\beta}}
\sigma_{\vec{r}+\vec{\beta}}, \label{Eul}
\ee
and the corresponding vertices have the form
$\chi_{8}= (2+2k)/3,~~~\chi_{9}=
(4+4k)/3,~~~\chi_{10}=
(-1+2k)/3,~~~\chi_{11}=(9+4k)/3,~~~\chi_{12}=
(4+4k)/3,~~~\chi_{13}=8+4k$
and $\chi_{1-7}$ are the same.
The limit $k \rightarrow \infty $ completely excludes the vertices
with self-intersections from the partition function.

{\Large{\bf 3.3}} Absolute value of the weights for vertex curvature.
The model with the absolute value of the Euler character
(\ref{Eulerchag})
\be
\Theta(M_2)=\sum_{<i,j>} \vert 2\pi - \alpha_{ij}-\beta_{ij}-...\vert
\ee
can be constructed if we take in (\ref{coupling}) and
(\ref{weight}) $\rho = 1$, and $g=-h=-1/4$ then
\be
a=-\frac{1}{2},~~~b=0,~~~c=d=-\frac{5}{12},~~~f=
\frac{1}{3},~~~e=0,~~~g=-h=-\frac{1}{4} \label{absolute}
\ee
with the corresponding weights
$$
\theta_{1} = 1,~~~\theta_{2}=0,~~~\theta_{3}=1,~~~\theta_{4}
= 0,~~~\theta_{5}= \theta_{6}=\theta_{7}= 2,$$
and self-intersection vertices are $\theta_{8}= (-4+2k)/3,~~~\theta_{9}=
(-8+4k)/3,~~~\theta_{10}= (5+2k)/3,~~~\theta_{11}=
(-15+4k)/3,~~~\theta_{12}=
(-8+4k)/3,~~~\theta_{13}=-12+4k.$

{\Large{\bf 3.4}} Increasing the dimension of the
lattice by one and leaving
the Hamiltonian (\ref{Euler}),(\ref{Eul}) and (\ref{abcdf})
(\ref{absolute}) without
changes one can see that this prime
system "moves"  to the next, physical, member
of the hierarchy and
describes now the  quantum gravity or more exactly the system
of fluctuating three dimensional manifolds with linear-gravity
action $A(M_3)$ (\ref{Hilquan}) on four dimensional lattice,

\be
Z_{Gravity}(\beta)=\sum_{ \{ \sigma \}}
exp( - \beta H^{3d}_{Euler} )~~~~~=~~~~~ \sum_{ \{ M_3 \}}
exp( - 2\beta A(M_{3}) ) \label{gravity}
\ee
where the Hamiltonian $H^{3d}_{Euler}$ exactly coincide with
(\ref{Euler}),(\ref{Eul}) and (\ref{abcdf}), (\ref{absolute}),
but the summation over
$\vec r$,$\vec \alpha$....is extended now over four dimensional lattice.

As we already have seen in (\ref{statsuper}),
(\ref{DiGrav}) and  (\ref{FeyGrav}), the  partition
function (\ref{gravity}) of the three-dimensional quantum
gravity can be
represented as a superposition of the random surfaces with
Euler action (\ref{statsuper}) or on the lattice
as the superposition of
linear string (\ref{DiGrav}) and  (\ref{FeyGrav}).

We can conclude from this that the spin system (\ref{gravity}),
which simulates three-dimensional  gravity undergoes the
second order phase transition in four dimensions
and that this phase transition should be
of the same nature as in 2d Ising ferromagnet or in
3d gonihedric system \cite{savvidy}. This happens
because they are the next to primary systems in the
geometrical hierarchy, that is they are next to
1d Ising, 2d gonimetric and 3d Euler systems
correspondingly.
Increasing the dimension of the lattice by one more unit we
will describe finally four-dimensional Gravity with area action
$S(M_{4})$ which is embedded into five dimensional lattice.

{\Large{\bf 4.}}In this paper we  clarify  the
point that the gonihedric string is the $superposition$ of the
weakly interacting fermions. We extend this result to quantum
gravity, which now appears as a superposition of weakly interacting
gonihedric strings.  We have proposed also an
alternative linear action $A(M_{4})$ for the four and higher
dimensional
quantum gravity.

We would like to thank L.Alvarez-Gaume, D.Gross, H.Nielsen,
A.M.Polyakov, F.Wegner and R.Schneider for discussions and support.
One of the authors (G.K.S.) acknowledges, that part of this
work was supported by the Alexander von Humboldt
Foundation and thanks Pavel Savvidy for his help in
computations.

\end{document}